\documentclass[final,floatfix,twocolumn,pre,aps,floats,amsfonts,amssymb,showpacs]{revtex4}

\usepackage{graphicx}
\usepackage{bm}

\def\EQ{\begin{equation}}
\def\EN{\end{equation}}
\def\EQA{\begin{eqnarray}}
\def\ENA{\end{eqnarray}}

\begin{document}
\title{  
Scaling in large Prandtl number turbulent thermal convection}
\author{ B. Dubrulle$^{1}$}

\address{
$^1$ CNRS, Groupe Instabilit\'e et Turbulence, CEA/DSM/DRECAM/SPEC, F-91191 Gif sur Yvette Cedex, France}

\begin{abstract}
We study the scaling properties of heat transfer $Nu$ in turbulent thermal convection at large Prandtl number $Pr$ using a quasi-linear theory. We show that two regimes arise, depending on the Reynolds number $Re$. At low Reynolds number, $Nu Pr^{-1/2}$ and $Re$ are a function of $Ra Pr^{-3/2}$. At large Reynolds number $Nu Pr^{1/3}$ and $Re Pr$ are function only of $Ra Pr^{2/3}$ (within logarithmic corrections). In practice, since $Nu$ is always close to $Ra^{1/3}$, this corresponds to a much weaker dependence of the heat transfer in the Prandtl number at low Reynolds number than at large Reynolds number. This difference may solve an existing controversy between measurements in SF6 (large $Re$) and in alcohol/water (lower $Re$). We link these regimes with a possible global bifurcation in the turbulent mean flow. We further show how a scaling theory could be used to describe these two regimes through a single universal function. This function presents a bimodal character for intermediate range of Reynolds number. We explain this bimodality in term of two dissipation regimes, one in which fluctuation dominate, and one in which mean flow dominates. Altogether, our results provide a six parameters fit of the curve $Nu(Ra,Pr)$ which may be used to describe all measurements at $Pr\ge 0.7$.
 \end{abstract}

\pacs{47.27 -i Turbulent flows, convection and heat transfer - 
47.27.Eq Turbulence simulation and modeling -
47.27.Te Convection and heat transfer }

\date{Eur. Phys. J. B vol 28 p 361-367 (2002)}

\maketitle

\section{Motivation and objectives}
In September 2001, a workshop on ``high Rayleigh number convection" was held at Illmenau (Germany). During the course of this workshop, a controversy arose about the scaling and magnitude of the heat transport, in large Prandtl number experiments. Two groups, one in Santa Barbara \cite{XBA00}, and one in Hong Kong \cite{XLZ02,LSZX02}, reported measurements performed in water and organic fluids (alcohols) at Prandtl numbers ranging approximately $Pr=4$ to $Pr=1300$ and in cell of aspect ratio unity. In these experiments, the non-dimensional heat transfer between the bottom and top, $Nu$, appears to depend only weakly on the Prandtl number, and to increase with the Rayleigh number $Ra$, like: 
\EQ
Nu=0.3 Ra^{0.28} Pr^{-0.028}.
\label{nusbhk}
\EN
A third group of people, located in Rehovot, reported measurements performed with $SF_6$ gas, near critical point. By varying the distance to critical point, one can vary the Prandtl number between $Pr=1.4$ up to $Pr=36$, with relatively small non-Boussinesq effects up to $Ra=10^{15}$ \cite{AshkStein99a}. The heat transfer measurements in these experiments revealed a much stronger Prandtl number dependence, and appeared to be about 1.5 larger in magnitude than in the water/alcohol experiments in the same range of Rayleigh numbers:
\EQ
Nu=0.51 Ra^{0.29} Pr^{-0.15}.
\label{nuasst}
\EN
Since the aspect ratio in the $SF_6$ experiments was comparable ($0.72$ v.s. $1$) to the water/alcohol experiments, there was no clear explanation of the discrepancy between the two type of measurements. Compressibility effects due to the vicinity of the critical point were mentioned as a possibility, even though simple estimates seem to rule out their influence in the range of Rayleigh number explored \cite{AshkStein99a}.\

Since then, the Rehovot group has improved its method of determination of the Nusselt number, which previously gave only precise estimate of {\sl relative} changes, not absolute ones. The improvement results in a better determination of the prefactor of the law (\ref{nuasst}) which appears to be decreased by a factor $2$ with respect to the old measurements (Steinberg, private communication). This solves the discrepancy in magnitude between the Nusselt measurents in alcohol/water and in SF6. But the discrepancy in the Prandtl dependence still remains.\

Clearly, in this type of controversy, it would be helpful to rely on theoretical results to guide our intuition. The water/alcohol experiments appear to be in good agreement with a recent theory by Grossmann and Lohse \cite{GL00,GL01}, which predicts a very weak decrease of the Nusselt number with the Prandtl number, due to a saturation of the viscous boundary layer (it increases with $Pr$). On the other hand, the $SF_6$ experiment is in excellent agreement with a theory by Shraiman and Siggia \cite{ShraSigg90}, based on the existence of a turbulent boundary layer. Since both theories are mutually excluding, they can however not be used to solve the discrepancy. In a recent work \cite{DLS02,Dubr00,Dubr01}, we have developed a quasi-linear theory of turbulent convection, which enables analytical predictions of the scaling laws of convection, including logarithmic corrections. The $SF_6$ experiment appeared to fit nicely into one scaling regime, in which the energy dissipation is dominated by the fluctuations, while the heat dissipation is dominated by the mean flow. The theory of Shraiman and Siggia also appeared as a special case of this theory, in which both the energy and the temperature dissipation are dominated by mean quantities. So there are actually two regimes which could explain the $SF_6$ experiment. It is then natural to wonder first which regime is the more suitable for explaining the $SF_6$ and whether the quasi-linear theory can be used to propose solutions to the water/alcohol-$SF_6$ controversy. This is the purpose of the present letter.

\section{The quasi-linear theory in short: ideas, weaknesses, constraints}

In this context, it is interesting to summarize the main ideas behind the quasi-linear theory. This theory is first used to compute analytically, from the Boussinesq equations, the mean and fluctuating velocity and heat profile within the turbulent boundary layer as a function of the Rayleigh and Prandtl numbers. Under the assumption that the energy and temperature dissipation are controlled by the turbulent boundary layer, these results can then be used to evaluate them analytically as a function of the Rayleigh and Prandtl number. Because this evaluation involves vertical integration of the local dissipation over the height of the turbulent boundary layer, the final result also becomes a function of the turbulent boundary layer scale. In the quasi-linear approximation, this scale can be computed by assuming that the velocity or thermal fluctuations are passively advected by the mean flow \cite{ShraSigg90,Dubr00}. However, because this mean flow a priori originates from the bulk (it is external to the turbulent boundary layer), it cannot be analytically computed and has to be taken as an external parameter. The number of degree of freedom can be greatly reduced if one assumes an algebraic profile $U=z^\epsilon$ for this mean flow. Indeed, by matching this mean flow with the analytic expression within the boundary layer, the parameter $\epsilon$ becomes the only free relevant parameter of the problem. In \cite{Dubr01}, we used velocity profiles given by numerical simulations by Kerr \cite{Kerr96,KerrHerr00} and Verzicco and Camussi \cite{VerzCamu99} to postulate that for $Pr<0.7$, the parameter $\epsilon=0$, while for $Pr>0.7$, its value is $\epsilon=-1/2$. This difference then provides a natural explanation of the difference between low and large Prandtl number measurements. Also, because bulk circulation can be severely affected by the geometry of the container (via the aspect ratio), it introduces a non-trivial dependence in the geometry which is difficult to quantify. In this regard, it is a weak point of the theory.\

On the other hand, we show below that the theory offers possibilities to connect different parameters of the convective cell independently of the bulk circulation, and depending only on the nature of the process which drives the dissipation (i.e. whether it is the mean flow or the fluctuations). Also, 
the hypothesis about the bulk flow contains several possibilities of direct check of the theory, via comparison between boundary layers length scales, and velocity profiles. Indeed, it can be shown \cite{Dubr01} that the exponent of the velocity profile and the length scale of the boundary layers are related through:
\EQA
\lambda_{BLT}&=&Pr^{\epsilon/(2+\epsilon)}u_\tau^{-(1+\epsilon)/(2+\epsilon)}\left(\frac{1}{\ln\lambda_{BLT}}+\frac{1}{f}\right)^{\delta(\epsilon)/(2+\epsilon)},\nonumber\\
\lambda_{BLV}&=&\left(\frac{Pr}{ u_\tau}\right)^{(1+\epsilon)/(2+\epsilon)}\left(\frac{1}{\ln\lambda_{BLV}}+\frac{1}{f}\right)^{\delta(\epsilon)/(2+\epsilon)}.
\label{resume}
\ENA
Here, $u_\tau= \sqrt{-\nu \partial_z U_x}$ is the friction velocity, and $\delta(\epsilon)$ is equal to 1 if $\epsilon=0$ and zero elsewhere. All quantities have been non-dimensionalized by the vertical gap and the thermal diffusivity. The subscript $T$ stands for thermal, and $V$ stands for velocity boundary layer.
The dependence of $\epsilon$ in $\lambda_{BLT}$ and $\lambda_{BLV}$ has been omitted for simpler notations. Also, the logarithmic and constant velocity regime have been lumped into the single notation $U(\lambda)\sim u_\tau/(1/\ln\lambda+1/f)$ which patches the two consecutive characteristic behaviors.
Note that in these units, $u_\tau/Pr=Re_\ast$, the Reynolds number based on the friction velocity\footnote{The subscript $\ast$ has been added to differentiate this Reynolds number from the Reynolds number $Re$ based on the large scale circulation, which will be introduced afterwards, and which is usually used in the scaling predictions.}. Equation (\ref{resume}) then encompasses  well-known relations. For $\epsilon=0$, one finds $\lambda_{BLV}\sim Re_\ast^{-1/2}$, like in a typical (logarithmic, i.e. $\epsilon=0$) boundary layer. For $\epsilon=1$, like in a laminar layer, one finds $\lambda_{BLV}\sim Re_\ast^{-2/3}$. This is the scaling originally proposed by Shraiman and Siggia \cite{ShraSigg90}.\

\section{Application to large Prandtl number convection}
\subsection{Reynolds number}
What would be the situation in large Prandtl number convection? Because $\nu\gg 1$, it can be expected that small-scale temperature fluctuations, if any, are mainly affected by the laminar part of the velocity field. Applying (\ref{resume}) with $\epsilon=1$, one then expect that the thermal boundary layer scales like $\lambda_{BLT}\sim Pr^{1/3} u_\tau^{-2/3}$. If the thermal fluctuations are so weak that only the mean temperature dominates the dissipation, then $\lambda_{BLT}\sim 1/Nu$ \cite{Dubr01}, fixing the link between $Re_\ast$ and $Nu$ as:
\EQ
Re_\ast\sim Nu^{3/2} Pr^{-1/2}.
\label{general}
\EN
These laws are quite general, and independent of the shape of the velocity in the bulk region. We thus expect them to be universal laws of the large Prandtl number regime. So, is it satisfied by the available measurements in both water/alcohol and $SF_6$?\

In the SF6 case, there are no direct measurements available of the friction velocity Reynolds number $Re_\ast$. Instead, Ashkenazi and Steinberg \cite{AshkStein99a} provide measurements of a Reynolds number based on the peak frequency $f_p$ from the velocity power spectra $Re=4f_p L^2/\nu$. It can be fitted by a power-law form $Re=2.6 Ra^{0.43\pm0.02} Pr^{-0.74\pm 0.02}$. Comparing the power-law form of $Nu$ and $Re$, one then sees that the law (\ref{general}) is very well satisfied provided $Re\sim Re_\ast$, as already stressed in \cite{AshkStein99a}.\

In the alcohol case, the Hong-Kong group measured $Nu=Ra^{0.278} Pr^{-0.028}$. Other  direct measurements concern not the friction Reynolds number, but the large scale Reynolds number $Re$. Its variation with $Pr$ and $Ra$ has been summarized by the Hong Kong group as: $Re=1.09 Ra^{0.43} Pr^{-0.76}$ \cite{LSZX02}. Interestingly enough, this variation is very similar to the variation deduced in SF6 (with a smaller prefactor, resulting in smaller Reynolds numbers). However, given the very weak dependence of $Nu$ with $Pr$, this measurements is clearly incompatible with (\ref{general}), which shows that in the alcohol case, either the equality $Re\sim Re_\ast$ does not hold, or/and the general law (\ref{general}) is violated. In that respect, it is interesting to note that a direct measurements of $Re_\ast$ performed by the same group, in a water experiment at $Pr=7$ for $Ra$ between $10^7$ to $10^{11}$ \cite{XinXia97} provides additional support of the violations of this laws. Indeed, they found $Re\sim Ra^{0.5}$, while $Re_\ast\sim Ra^{0.33}$, closer to $Re_\ast\sim Nu\sim Re^{2/3}$ than to (\ref{general}). These interesting new scalings are reminiscent of measurements performed on the r.m.s. vertical velocity fluctuations by the Rehovot group, which indicate that the corresponding Reynolds number obeys a scaling transition from $Ra^{0.43\pm 0.02}$  at $Pr=27,45$ and $93$, to a scaling $Ra^{0.34}$ at $Pr=190$ \cite{AshkStein99b}.\

The alcohol measurements show that our initial naive expectation gathered in (\ref{general}) has to be refined. For this, let us come back to the main hypothesis governing (\ref{general}). It is based on the idea that the thermal boundary layer is determined by the balance between the large scale advection of temperature fluctuations $U\partial \theta$ and the heat dissipation $\partial^2 \theta\sim \theta Nu^{-2}$. Within the boundary layer, the flow is laminar, oriented in the longitudinal direction and obeys the friction law $U=Pr Re_\ast^2 z$, where $z$ is the vertical coordinate \cite{Dubr01}. To obtain a "typical" velocity advection, one then sets $z=\lambda_{BLT}=Pr^{-1/3} Re_\ast^{-2/3}$ to get $Re_\ast^2 Pr =Nu^3$. This argument does not take into account the constraint that the typical size of the boundary layer cannot increase beyond the size of the box \cite{GL01}. This effect is obtained when the whole cell is laminar, i.e. as the Reynolds number decreases. To take this into account, we may then change our estimate into $U=Pr Re_\ast^2/(A+Pr^{1/3}Re_\ast^{2/3})$, resulting into the more general law:
\EQ
\frac{Re_\ast}{\sqrt{1+c_0 Pr^{1/3}Re_\ast^{2/3}}}=\frac{Nu}{Pr^{1/2}},
\label{general2}
\EN
where $c_0$ is a constant depending a priori on the aspect ratio. Decreasing $Re_\ast$ (e.g. by increasing $Pr$ as in the experiment of Ashkenazi and Steinberg), one observes a transition from $Re_\ast\sim Nu^{3/2} Pr^{-1/2}$ (i.e. the naive law (\ref{general})) to $Re_\ast \sim Nu Pr^{-1/2}$.\

\begin{figure}[hhh]
\includegraphics[clip=true,width=0.99\columnwidth]{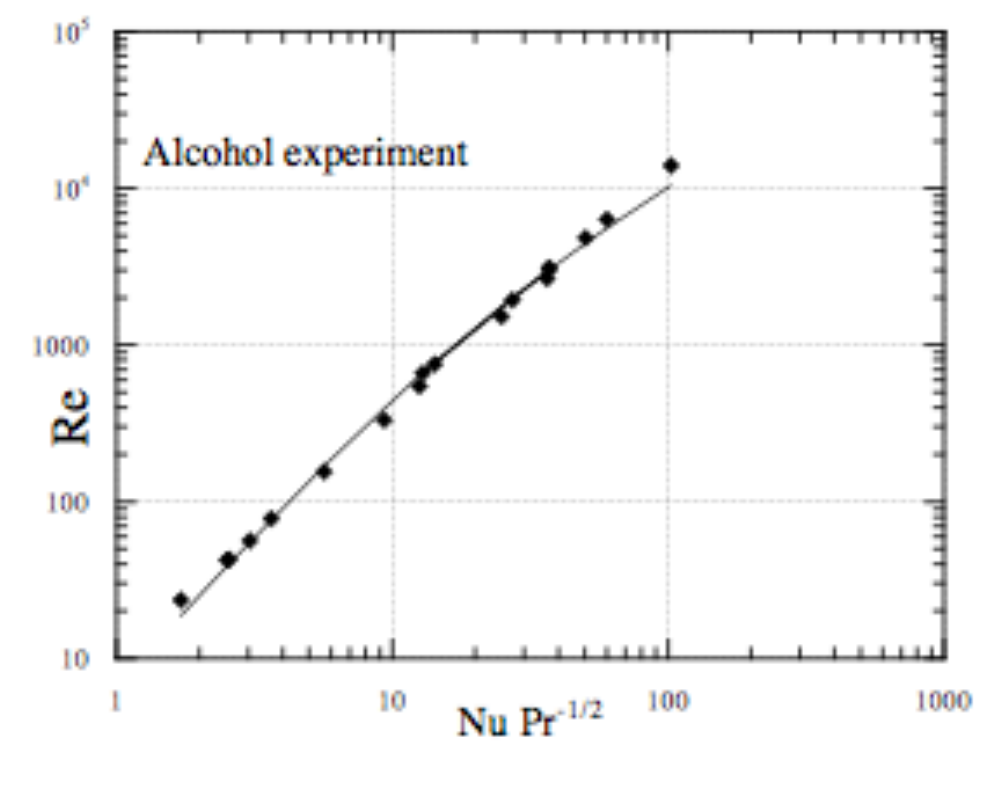}
\caption[]{ Reynolds number versus Nusselt number in the alcohol experiment \protect\cite{XLZ02}). The symbols are the experimental measurements. The line is the fit predicted by the model (eq. (\ref{reynoldsun}) with $c_1=7$ and $Re_c=17$).}
\label{fig:fig1}
\end{figure}

\begin{figure}[hhh]
\includegraphics[clip=true,width=0.99\columnwidth]{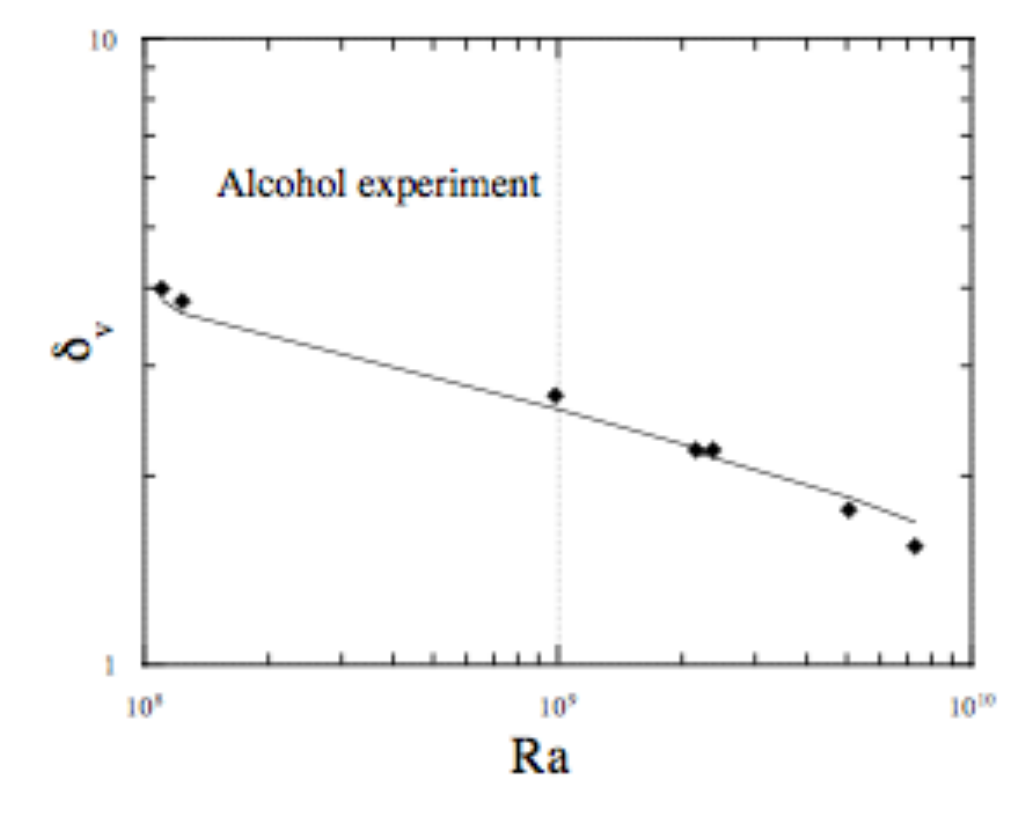}
\caption[]{Boundary layer scale in the alcohol experiment at $3<Pr<10$). The symbols are the model prediction $\delta_v=1/(1+Re/17)$. The line is the fit $\delta_v=Ra^{-0.16} Pr^{0.24}$ proposed by \protect\cite{XLZ02}.}
\label{fig:fig2}
\end{figure}

\begin{figure}[hhh]
\includegraphics[clip=true,width=0.99\columnwidth]{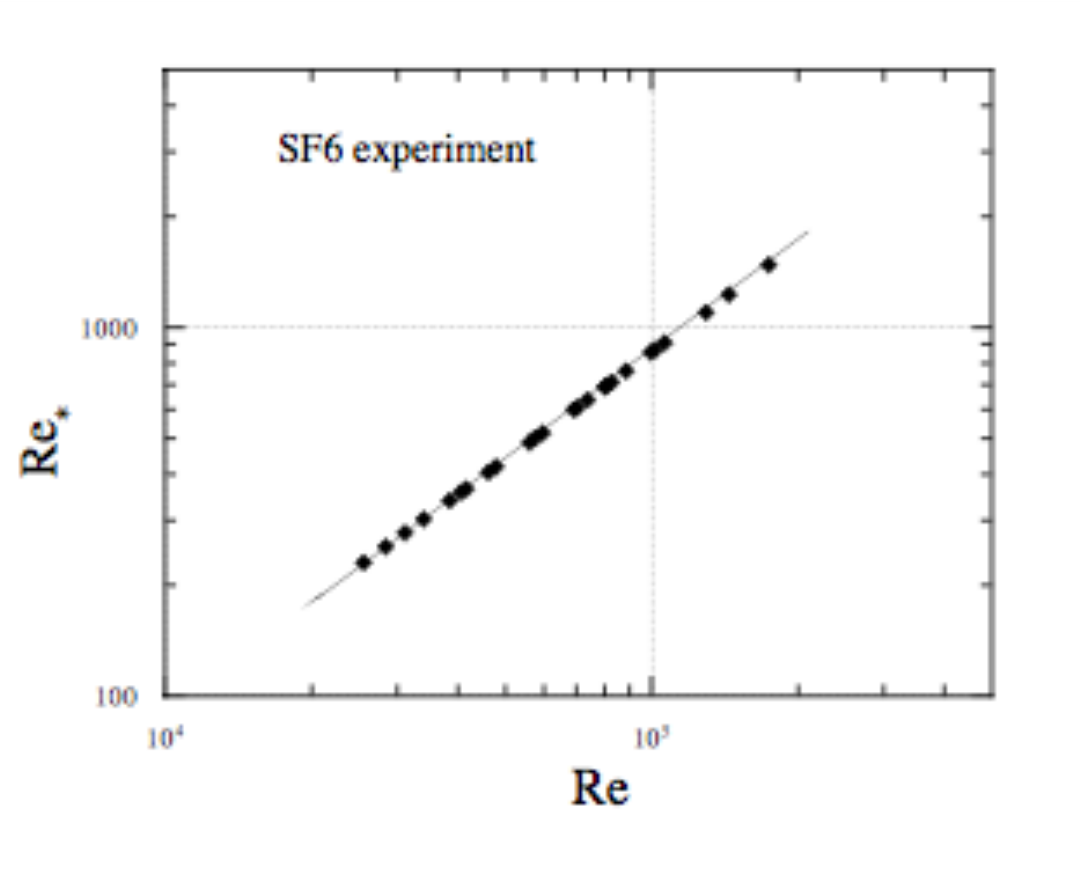}
\caption[]{Friction Reynolds number in the SF6 experiment. The symbols is the estimated friction Reynolds number derived using the theoretical formula (eq. (\ref{reynoldsun}) with $c_1=7$ and $Re_c=17$). The  line is a fit $Re_\ast\sim Re$, a behavior predicted by the theory.}
\label{fig:fig3}
\end{figure}

Large scale circulations are easier to measure than friction velocities. Its is therefore interesting to connect $Re$ and $Re_\ast$ to be able to exploit and verify (\ref{general2}). For this, we match the large scale circulation velocity $Re Pr$ to the velocity $v_{BL}$ at the top of the boundary layer. At large Reynolds number, the boundary layer is turbulent and $v_{BL}=Re_\ast Pr=Re Pr$ (within logarithmic corrections). At smaller Reynolds number, the boundary layer is laminar and $v_{BL}=Re_\ast^2 \lambda_{BLT}$. The link between $Re$ and $Re_\ast$ then depends on the shape of the velocity in the bulk flow (through the value of $\epsilon$). However, we may note that this length-scale is also subject to the condition that it cannot increase beyond the size of the box \cite{GL01} (a situation occurring at very low Reynolds numbers). This remarks allows us to build a general law resulting from a patching between the large Reynolds number case and the very low Reynolds number case as:
\EQ
Re=c_1\frac{Re_\ast^2}{1+Re^\ast/Re_c},
\label{reynoldsun}
\EN
where $c_1$ is a non-universal constant and $Re_c$ is a critical Reynolds number. A rough estimate of this two parameters can be made using the alcohol data of Xia and collaborators. The available measurements unfortunately only concern $Nu$ and $Re$ as a function of $Ra$ and $Pr$. However, in this low Reynolds number case, we expect $R_\ast\approx Nu Pr^{-1/2}$. Using this, we may then find $c_1$ and $Re_c$ by a two parameter fit of $Re$ versus $Nu Pr^{-1/2}$.
Fig. \ref{fig:fig1} shows a comparison of the measured $Re$ and the fit using (\ref{reynoldsun}) with $c_1=7$, $Re_c=17$. The agreement is quite satisfactory. A further test of this fit can be made by comparing the effective boundary layer lentgh scale $\lambda_{BLV}\equiv Re/Re_\ast^2=1/(1+Re_\ast/Re_c)$ with the direct measurement of Xian and Xia \cite{XinXia97,LSZX02} in water ($Pr=7$) and alcohol, giving $\lambda_{BLV}\sim
Ra^{-0.16}Pr^{0.24}$. One sees in Fig. \ref{fig:fig2} that the agreement is satisfactory. Finally, one can invert (\ref{reynoldsun}) to obtain $Re_\ast$ as a function of $Re$ in the SF6 experiment, using the values of the parameters calibrated from the alcohol experiment. The results is shown in Fig. \ref{fig:fig3}. It indeed give $Re\sim Re_\ast$, as expected in this large Reynolds number case.\

The combination of (\ref{reynoldsun}) and (\ref{general2}) predicts that, for a given (large) $Pr$, large Reynolds number leads to $Re_\ast=Re=Nu^{3/2}Pr^{-1/2}$, while at lower Reynolds number, $Re=Re_\ast^2=Nu^2/Pr$. In between, there is a transitional regime in which $Re$ varies smoothly with $Re_\ast$. For the range of Reynolds numbers explored by the Hong-Kong group, we checked that this corresponds to an apparent scaling behavior $Re=Re_\ast^{2/3}=Nu^{3/2} Pr^{-3/4}$. The first case is representative of the data by Ashkenazi and Steinberg, while the second is representative of the data by Xia and collaborators, and Ahlers and collaborators. It seems therefore that we have identified a possible cause of discrepancy between the two groups: the difference in magnitude in the Reynolds number, resulting in a different regime explored. This difference of Reynolds number, observed in experiments with similar aspect ratios, Prandtl numbers and Rayleigh numbers is surprising at first sight. A first explanation is that compressibility effects do play a role. A second explanation is that for this range of parameters, there are two possible solutions, with distinct transport properties (leading to different large-scale circulations and different Nusselt numbers). Note that a similar phenomenon has recently be observed in the flow between two coaxial rotating cylinders (the von Karman flow). In a certain range of Reynolds number, the mean component of the turbulent flow undergoes a global bifurcation with three distinct flow configurations, characterized by different transport properties, and subject to a strong hysteresis \cite{Marie01}. Is one of these two explanations sufficient to explain the difference in Prandtl number dependence observed? To answer this question, we now focused on the second ingredient of \cite{ShraSigg90,GL00,Dubr01}: the energy dissipation, enabling to link $Re$, and $Ra$ and $Nu$.

\subsection{Dissipation}
 In \cite{Dubr01}, we have shown that the contribution to the energy dissipation scales like $u_\tau^4 \lambda_{BLV}/Pr=Pr^3 Re_\ast^4\lambda_{BLV}$ for a laminar mean flow,
$u_\tau^3=Pr^3 Re_\ast^3$ for a turbulent mean flow (larger Reynolds number), and like $(Pr Nu)^{3/2} Ra^{1/2}
ln \lambda_{BLV}/\lambda_{BLV}^2 $ for fluctuations. The first two contributions can be patched together as a function of the Reynolds number in a way similar to what we did for the Reynolds number. Summing the contributions of the mean flow and of the fluctuations (in a way similar to a procedure adopted by Grossmann and Lohse \cite{GL00,GL01}) and recalling that this energy dissipation equals $Pr Ra (Nu-1)$ \cite{ShraSigg90}, we obtain in this framework a general relation linking $Re_\ast$ and the other parameters:
\EQ
Pr Ra (Nu-1)=c_1 Pr^3 \frac{Re_\ast^4}{1+Re_\ast/Re_c}+c_2 (Pr Nu)^{3/2} Ra^{1/2}
ln \lambda_{BLV}/\lambda_{BLV}^2,
\label{dissipation}
\EN
where $c_1$ and $c_2$ are (a priori non universal) constants. Using (\ref{general2}), this relation can also be written in a more illuminating form as:
\EQ
\frac{Ra}{Pr^{3/2}}= c_1 \frac{Nu}{Pr^{1/2}}Re_\ast^2 \frac{1+c_0 Pr^{1/3} Re_\ast^{2/3}}{1+ Re_\ast/Re_c}+c_2 \frac{Nu^{1/2}}{Pr^{1/4}} \frac{Ra^{1/2}}{Pr^{3/4}}
ln \lambda_{BLV}/\lambda_{BLV}^2.
\label{dissipationbo}
\EN
This relation can be used to determine interesting scaling properties independent of the bulk flows or of the geometry. Indeed, we note that $\lambda_{BLV}$ is a function of $Re_\ast$ (see (\ref{resume})). Further, we see from (\ref{general2}) that at low Reynolds number $Re_\ast$ is a function only of $Nu Pr^{-1/2}$. Considering (\ref{dissipationbo}) at low Reynolds, this means that in this regime $Nu Pr^{-1/2}$ and $Re$ are  function only of $Ra Pr^{-3/2}$. This conclusion is independent of the fact that the dissipation is dominated by the mean flow or by the fluctuations. In practice, since $Nu$ is close to $Ra^{1/3}$ and $Re$ close to $Ra^{1/2}$, this corresponds to a very weak dependence of $Nu$ with $Pr$, and a dependence $Pr^{-0.75}$ for $Re$.\

At large Reynolds number, $Re_\ast$ is a function of $Nu Pr^{-1/3}$ and $\lambda_{BLV}=Re_\ast^{1/3}$ \cite{Dubr01}. Inserting this into (\ref{dissipationbo}), we see that at large Reynolds number, $Re_\ast Pr$ is a function only of $Ra Pr^{-1/3}$ (within logarithmic corrections), or equivalently $Nu Pr^{1/3}$ is a function only of $Ra Pr^{2/3}$. For $Nu$ and $Re$ close to a $Ra^{1/3}$ and $Ra^{1/2}$ dependence, this means a decrease of $Nu$ like $Pr^{-1/9}$ (within logarithmic corrections) and of $Re$ like $Pr^{-2/3}$. The decrease in $Nu$ in that regime is therefore stronger than in the low Reynolds number regime, and qualitatively explains the difference between the experiments in SF6 and in alcohol/water. For a quantitative agreement, one would need to evaluate the constants $c_1$,..$c_2$ appearing in (\ref{dissipationbo}) and (\ref{general2}). One possibility would be to perform a non-linear fit of the experimental data points, like \cite{GL01}. In the sequel, we would like to explore another possibility, based on a collapse method inspired from finite size scaling in statistical mechanics.

\subsection{A scaling approach}

Equation (\ref{dissipationbo}), valid in both the large and low Reynolds number regime, provides an interesting implicit equation for the variable $y=\sqrt{Ra/(Nu Pr)}$:
\EQ
y^2= c_1 Re_\ast^2 \frac{1+c_0 Pr^{1/3} Re_\ast^{2/3}}{1+ Re_\ast/Re_c}+c_2 y
ln \lambda_{BLV}/\lambda_{BLV}^2.
\label{dissipationbo2}
\EN
The function $y(R_\ast,Pr)$ can be found once the functional shape for $\lambda_{BLV}$ is provided. In the sequel, we shall adopt 
\EQ
\lambda_{BLV}^{-1}=1+c_3 R_\ast^{1/3},
\label{patchlambda}
\EN
 patching between the large Reynolds number case, and the saturation expected at very low Reynolds number.\

Our goal is now to determine the constants $c_1,c_2,c_3$ and $Re_c$ from the available data. For this, we remark that as $P\to\infty$, $Re_\ast\to 0$ and the asymptotic behavior of $y$ is:
\EQ
y^2=c_1 Re_\ast^2,\quad Pr\to\infty.
\label{asymptotic}
\EN
Note that given (\ref{general}), this corresponds to $Nu\sim Ra^{1/3}$, a limiting law also found by (\cite{GL01}). In addition, experimental data indicate that in the range of parameters usually explored, the behavior of $Nu$ with $Ra$ and $Pr$ follows approximate power laws. This suggests that it should be possible to approximate the behavior of $y$ at large but finite Prandtl number via a scaling shape (reminiscent of finite size scaling in statistical mechanics):
\EQ
y^2=c_1 Re_\ast^2 F(Re_\ast Pr^\alpha),
\label{fsize}
\EN
where $\alpha$ is a scaling exponent (equivalent to a critical exponent) and $F$ is a (universal) scaling function. In this approximation, the two parameters ($\alpha$ and $F$) can be found experimentally from a collapse procedure on the data taken at different $Pr$ and $Re_\ast$. This procedure is described below. We stress that since the scaling form is not exactly satisfied by the theoretical formula (\ref{dissipationbo}), it will only provide an approximate experimental verification of (\ref{dissipationbo2}). However, since this method uses data in different experimental set ups (different $Pr$ and $Re_\ast$), it is less liable to systematic experimental errors and thus, provides a more robust estimate of the constants $c_1,...c_3$ than via a direct fit of the data. In addition, it provides a strong independent test of the theoretical formula (\ref{dissipationbo2}) since it does not use its functional shape {\sl a priori}.

\subsection{A collapse method}

\begin{figure}[hhh]
\includegraphics[clip=true,width=0.99\columnwidth]{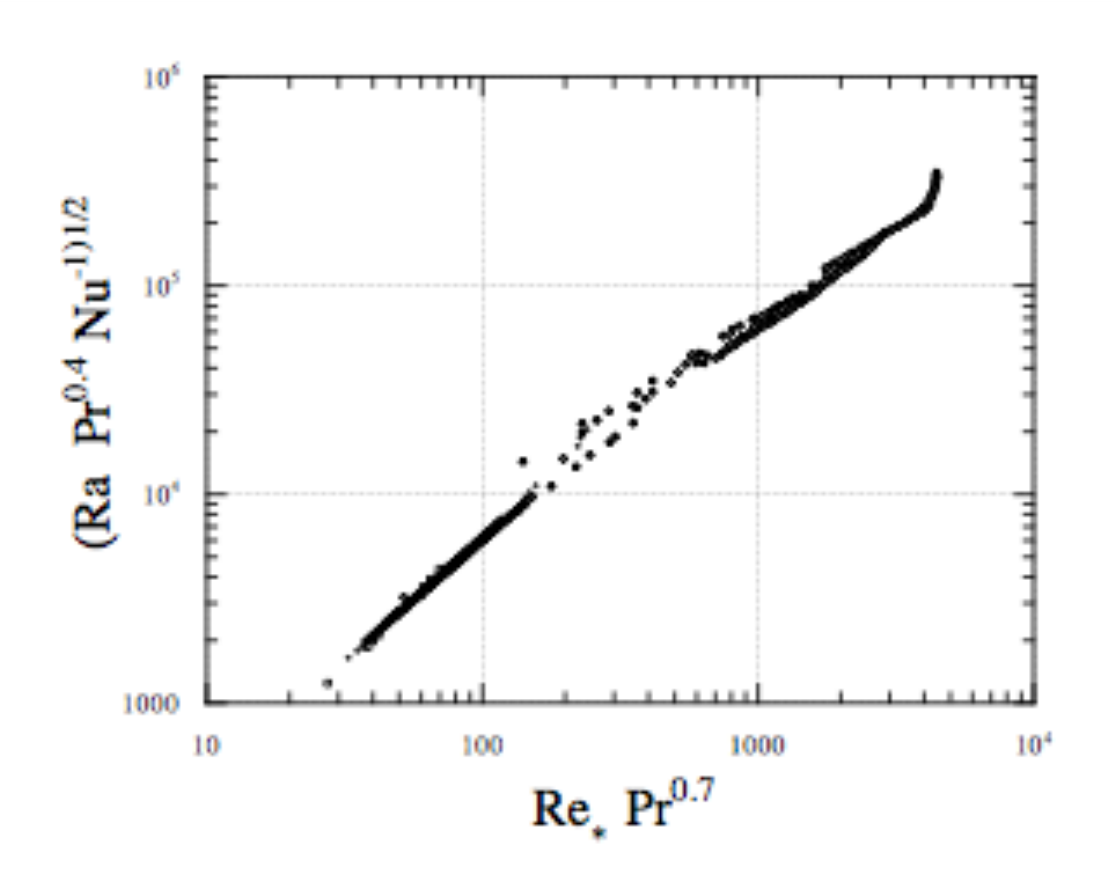}
\caption[]{ The universal scaling function. Best collapse of the data in alcohol \protect\cite{XLZ02}, in water \protect\cite{XBA00} and in SF6 \protect\cite{AshkStein99a}, obtained by plotting $\sqrt{Ra Pr^{0.4} Nu^{-1}}$ as a function of $Re_\ast Pr^0.7$. Note the bimodal regime appearing in intermediate range of $Re_\ast Pr^{0.7}$.}
\label{fig:fig4}
\end{figure}

\begin{figure}[hhh]
\includegraphics[clip=true,width=0.99\columnwidth]{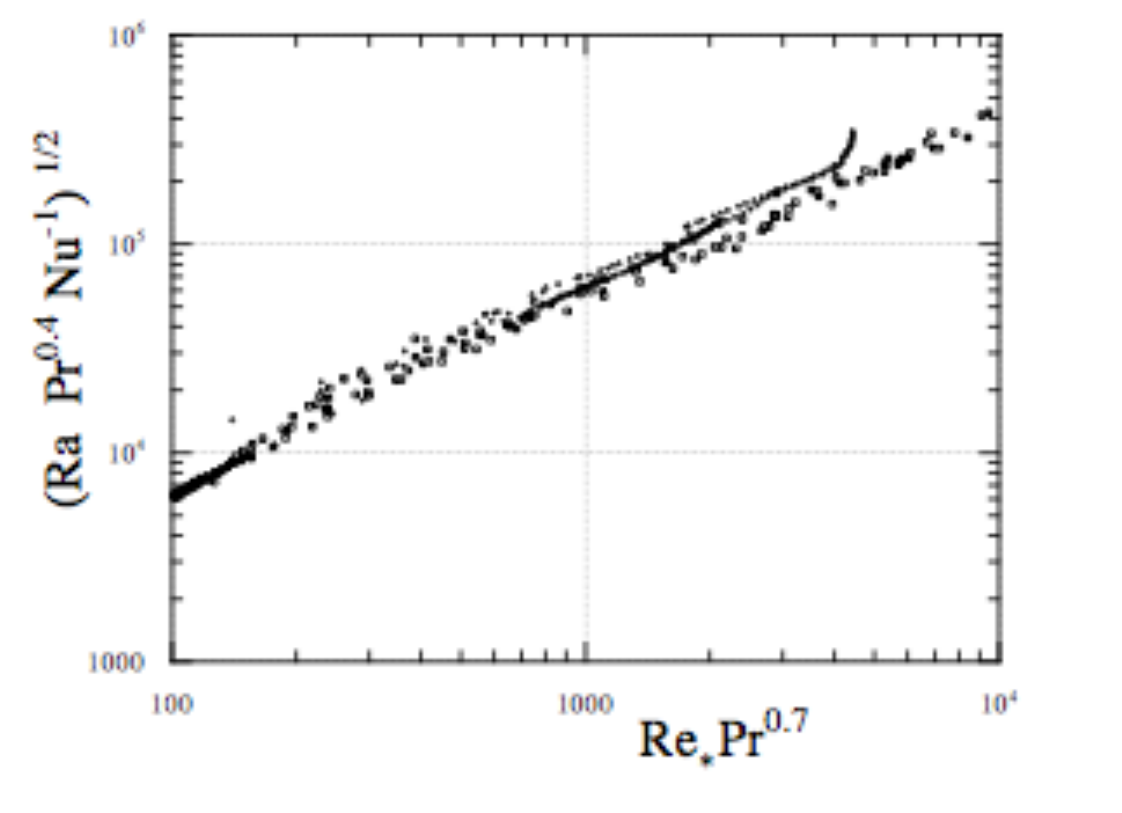}
\caption[]{Test of the bimodality. Same as Fig. 4, with the data of \protect\cite{chavanne} plotted as open symbols. The figure has been enlarged around the intermediate range of $Re_\ast Pr^{0.7}$ for more clarity. Notice the persistence of the bimodality, although it is slightly less apparent.}
\label{fig:fig5}
\end{figure}

In recent convection experiments, the quantity $Re_\ast$ is usually not directly measured and we have to use only an estimate of it, based on (\ref{general}). To minimize the error introduced by this procedure, we may rewrite (\ref{asymptotic}) as:
\EQ
y Pr^{\alpha}=G(Re_\ast Pr^\alpha),
\label{direct}
\EN
with $G(x)=\sqrt{c_1 x^2 F(x)}$. The determination of $\alpha$ and $G$ is then made by trial and error, plotting $y Pr^\alpha$ as a function of $Re_\ast Pr^\alpha$, and choosing $\alpha$ of that all curves, corresponding to different measurements at different $Pr$ and $Re$, collapse onto a single curve. The resulting curve is the scaling function $G$. Fig. \ref{fig:fig4} present the best collapse obtained with $\alpha=0.7$ for the data collected by the groups of Ahlers (Santa Barbara), Xia (Hong-Kong) and Steinberg (Rehovot) \footnote{We thank them warmly for making these data available to us}. In the first two cases, we used $Re_\ast=\sqrt{Nu^2/Pr}$. In the third case, we used $Re_\ast=0.17\sqrt{Nu^3/Pr}$. Interestingly enough, the scaling function seems to possess two branches at intermediate Reynolds number (one extending the Ahlers data, the other one extending the Steinberg data). Note that among the data of Steinberg and his group, part of the data do fall onto the lowest branch corresponding to the Ahlers data. To check whether  this is an artefact of our method, or whether it corresponds to a real physical process (like bimodality of the Nusselt numbers,  discussed in \cite{chavanne,CCCHCC97}, we ran an additional check. We present in Figure \ref{fig:fig5} the collapse obtained when the data of Chavanne et al \cite{chavanne} have been included, with the Reynolds number estimated as $Re_\ast=\sqrt{Nu^2/Pr}$. As can be seen, the universal scaling function seems also to describe these data. Moreover, the range where our bimodality was detected appears to coincide with the bimodality range of Chavanne et al. Such a bi-modality could arise for example from a global bifurcation of the turbulent mean flow, like in the von Karman experiment (see Section III.A.) 

\subsection{Fits and consistency check}

\begin{figure}[hhh]
\includegraphics[clip=true,width=0.99\columnwidth]{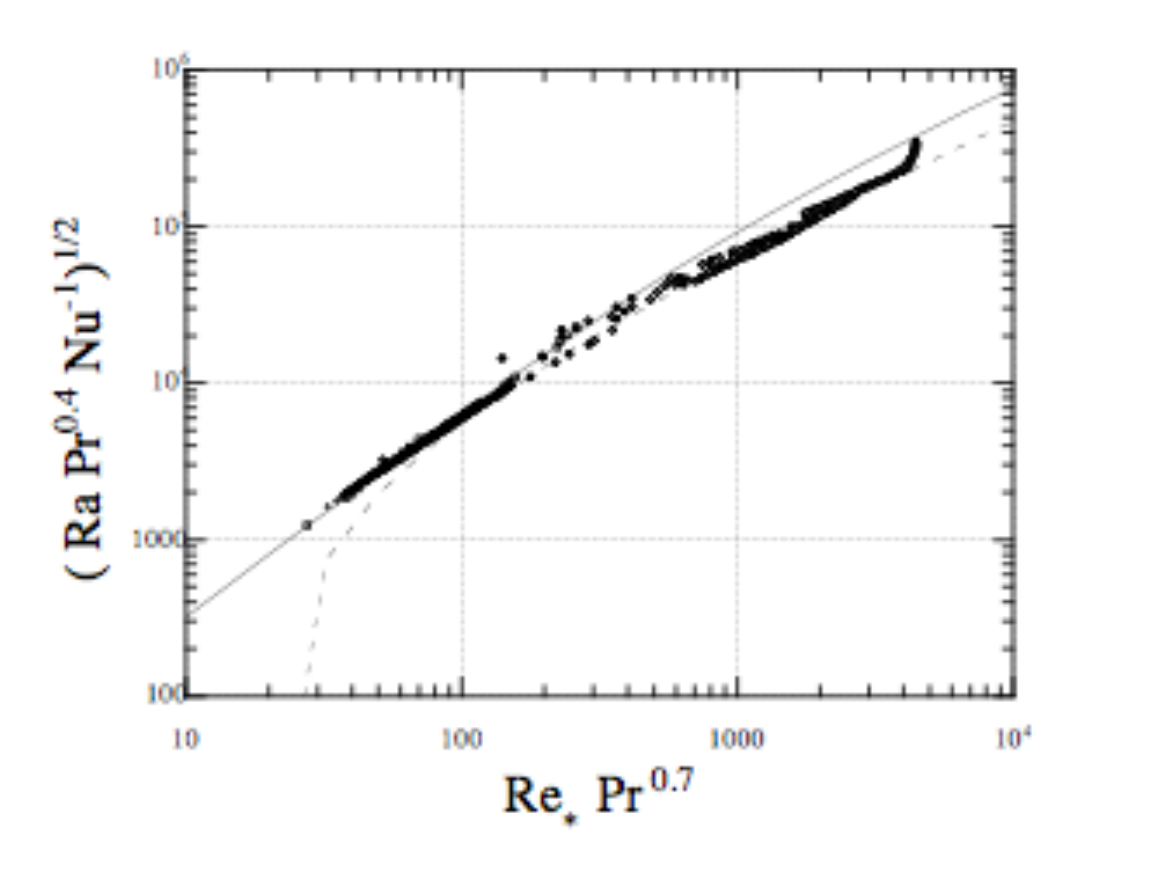}
\caption{ Fit of the universal function. The upper branch is fitted with formula 
(\ref{meanflowcontr}) (solid line), corresponding to a mean flow dominated dissipation. 
The "lower branch" is well fitted by a "fluctuation contribution"  formula
(\ref{fluctuations}), (dotted line).}
\label{fig:fig6}
\end{figure}

\begin{figure}[hhh]
\includegraphics[clip=true,width=0.99\columnwidth]{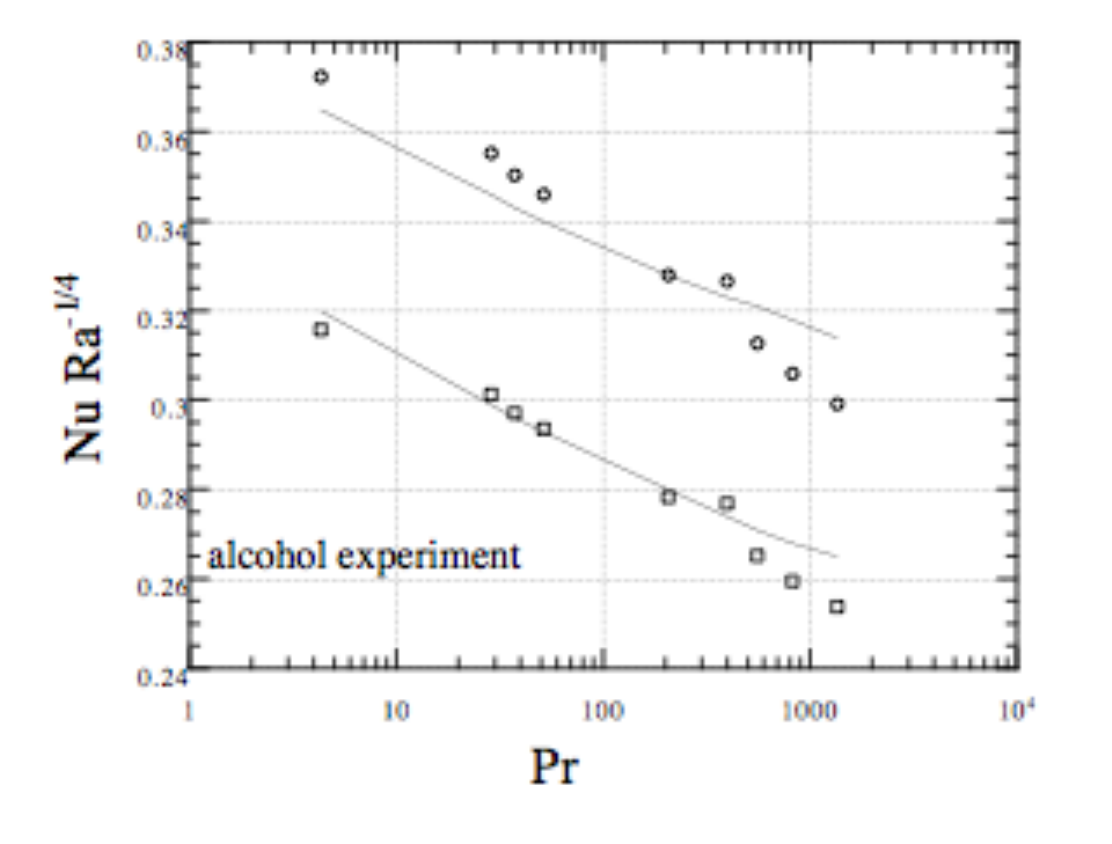}
\caption[]{ Nusselt versus Prandtl in the alcohol experiment. The open symbols are $Nu Ra^{-0.25}$ for $Ra=10^{9.25}$ (circles) and $Ra=10^{7.25}$ (squares). The line are the theoretical predictions, using the fit (\ref{meanflowcontr}).}
\label{fig:fig7}
\end{figure}

\begin{figure}[hhh]
\includegraphics[clip=true,width=0.99\columnwidth]{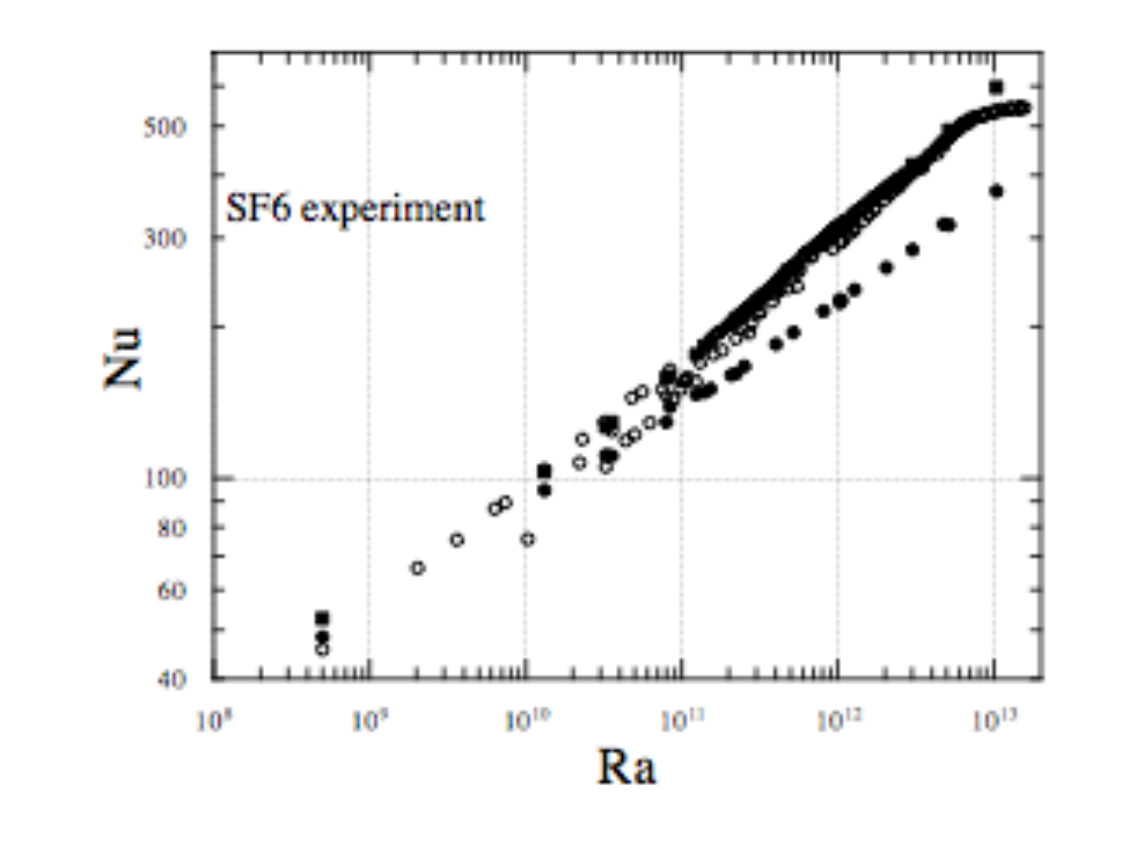}
\caption[]{ Nusselt versus Prandtl in the SF6 experiment. The open symbols are the experimental measurements. The filled symbols are the theoretical predictions, using the fit (\ref{meanflowcontr}) (circles) or (\ref{fluctuations}) (squares).}
\label{fig:fig8}
\end{figure}

This bimodality can be reproduced within the theoretical model developed in Section III.B. In this model, we showed that the dissipation is the sum of two contribution: one from the mean flow (term proportional to $c_1$), and one from the fluctuations (term proportional to $c_2$). We show on Figure \ref{fig:fig6} that the upper branch of the universal function is well fitted by a "mean flow contribution":
\EQ
y_{mf}=\sqrt{4.61 x^2 \frac{1+49.56 x^{2/3}}{1+x/600}},\quad x=Re_\ast Pr^{0.7}.
\label{meanflowcontr}
\EN
The "lower branch", on the other hand, is well fitted by a "fluctuation contribution" :
\EQ
y_{fl}=0.55 (1+30 x^{1/3})^2 \ln \left(0.012\left(1+30 x^{1/3}\right)\right).
\label{fluctuations}
\EN
The two corresponding regimes (fluctuation or mean flow dominated) seem to be mutually excluding, explaining the bimodality. We do not have an explanation for this. The quality of the fit can be checked {\sl a posteriori} by solving simultaneously (\ref{fluctuations}) or (\ref{meanflowcontr}) and (\ref{general}) for a given $Ra$ and $Pr$, to find $Nu(Ra,Pr)$. This consistency check is shown in Fig. \ref{fig:fig7} and \ref{fig:fig8} for the data of Ahlers, Xia and Steinberg. Overall, the quality of the fit is good. We expect improvement to be gained once direct measurements of $Re_\ast$ will be available.

\section{Summary}
We have shown that a discrepancy between heat flux in alcohol/water and in SF6, at large Prandtl numbers, could be explained by a Reynolds number effect and the existence of two different regimes, induced by a global bifurcation in the mean flow. We further showed how a scaling theory could be used to describe these two regimes through a single universal function. This function presents a bimodal character for intermediate range of Reynolds number. We explained this bimodality in term of two dissipation regimes, one in which fluctuation dominate, and one in which mean flow dominates. Altogether, our results provide a six parameters fit of the curve $Nu(Ra,Pr)$ which may be use to describe all measurements at $Pr\ge 0.7$. This is one more parameter than the theory recently developed by Grossmann and Lohse \cite{GL01}, but it also fits the data by Steinberg and it may lead to more precise estimates, as the fit is done simultaneously over different experiments.\

{\bf Aknowledgements} 
This work would not have been possible without the help of Victor Steinberg, Guenter Ahlers and Ke-Quing Xia who made their data and their latest results available to me. I thank them warmly for that.

\end{document}